\documentstyle[twocolumn,aas2pp4]{article}
\def\stacksymbols #1#2#3#4{\def\theguybelow{#2}
	\def\verticalposition{\lower#3pt}
	\def\spacingwithinsymbol{\baselineskip0pt\lineskip#4pt}
	\mathrel{\mathpalette\intermediary#1}}
\def\intermediary #1#2{\verticalposition\vbox{\spacingwithinsymbol
	\everycr={}\tabskip0pt
	\halign{$\mathsurround0pt#1\hfil##\hfil$\crcr#2\crcr
		\theguybelow\crcr}}}
\def\lta{\stacksymbols{<}{\sim}{2.5}{.2}}
\def\gta{\stacksymbols{>}{\sim}{3}{.5}}

\begin{document}
\title{RELATIVE SIZES OF X-RAY AND OPTICAL IMAGES OF 
ELLIPTICAL GALAXIES: CORRELATION WITH X-RAY LUMINOSITY$^1$}

\author{William G. Mathews$^2$ and Fabrizio Brighenti$^{2,3}$}

\affil{$^2$University of California Observatories/Lick Observatory,
Board of Studies in Astronomy and Astrophysics,
University of California, Santa Cruz, CA 95064\\
mathews@lick.ucsc.edu}

\affil{$^3$Dipartimento di Astronomia,
Universit\`a di Bologna,
via Zamboni 33,
Bologna 40126, Italy\\
brighenti@astbo3.bo.astro.it}






\vskip .2in

\begin{abstract}

Optical parameters of elliptial galaxies are 
tightly correlated, but their x-ray parameters vary widely.
The x-ray luminosity $L_x$ ranges over 
more than an order of magnitude for ellipticals 
having similar optical luminosity $L_B$.
The source of this scatter has been elusive.
We show here that the dispersion in $L_x$ for fixed 
optical luminosity $L_B$ correlates strongly 
with the dimensionless ratio of the 
sizes of the x-ray and optical images, $r_{ex}/r_e$.
A distance-independent variant of this correlation is
$L_x/L_B \propto (r_{ex}/r_e)^{0.60 \pm 0.30}$.
This correlation 
may be a natural result of mergings and tidal truncations 
that occurred during the early formation of 
ellipticals in groups of galaxies.
The shapes of the x-ray images also vary: some are
compact (e.g. NGC 4649, 7626, 5044), others diffuse
(e.g. NGC 4636, 1399).

\end{abstract}


\section{INTRODUCTION}

X-ray and optical luminosities of massive elliptical 
galaxies ($L_B \gta 1.6 \times 10^{43}$ erg s$^{-1}$) 
are correlated, 
$L_x \propto L_B^p$, with $p \approx 2.0 \pm 0.2$
(Eskridge, Fabbiano, \& Kim 1995a).
The exponent $p \approx 2$ 
can be explained as thermal x-ray emission from interstellar gas
heated to the galactic virial temperature $T \sim 10^7$ K
(e.g. Tsai \& Mathews 1995).
However, the scatter of $L_x$ 
about this correlation is enormous, considerably exceeding
an order of magnitude 
(e.g. Eskridge, Fabbiano, \& Kim 1995a).
Many attempts have been made to understand the reason for 
this large scatter but no single explanation is 
generally accepted
(D'Ercole et al. 1989; 
White \& Sarazin 1991; Ciotti et al. 1991; 
Mackie \& Fabbiano 1997).

In our recent studies of the gas-dynamical evolution of the 
hot interstellar gas in ellipticals,
we have found that most or all bright 
ellipticals contain an additional component of 
very old and extended hot gas in addition to 
gas ejected from currently evolving galactic stars 
(Brighenti \& Mathews 1997). 
This realization -- and the known variation in
the size of x-ray images (Loewenstein 1997) -- 
led us to speculate that the variation in $L_x$ 
could result from the transfer of hot gas and 
dark matter between galaxies 
when they resided in small, 
elliptical-rich groups.
Most of these groups observed today 
are dominated by a single large 
elliptical which has evidently grown in size and prominence
at the expense of other
group galaxies by some combination of mergers and tidal 
strippings.
These dynamical developments occur rapidly during the 
early phases of group formation
(Merritt 1985).
Some of these group-dominant ellipticals have now joined 
richer clusters and brought along their relative allotments
of hot halo gas.
With this evolutionary hypothesis in mind, we were led
to consider the relative sizes of the x-ray 
and optical images as a 
possible explanation for the large range in $L_x$ for given $L_B$.
In this letter we show that the relative sizes of the 
images does indeed 
correlate strongly with residuals 
measured from the $L_x \propto L_B^p$ relation.

\section{RATIO OF EFFECTIVE RADII AND $L_x$}

Optical sizes of elliptical galaxies are characterized 
by the effective radius $r_e$, defined as 
the radius that includes half the 
projected light.
Similarly we define an effective x-ray radius $r_{ex}$ that 
includes half the total x-ray luminosity $L_x$ in projection.
Unfortunately, relatively few elliptical galaxies have been 
observed with sufficient spatial resolution to allow 
a determination of $r_{ex}$ and, for those 
galaxies having sufficient resolution, 
$r_{ex}$ is not explicitly provided.
In a search through the recent literature we have identified 
eleven ellipticals having well resolved x-ray surface 
brightness distributions $\Sigma_x(r)$ and for which 
the local background radiation has been removed.
We have determined 
the effective radius $r_{ex}$ from the defining 
integral condition 
$$ \int_0^{r_{ex}} \Sigma_x(r) 2 \pi r dr 
= {1 \over 2} \int_0^{r_t} \Sigma_x(r) 2 \pi r dr $$
where $r_t$ is the outermost extent of the 
(background-subtracted) x-ray image.
In computing the integrals above we assume that $\Sigma_x$ 
is constant within the innermost observed radius.
Our values of $r_{ex}$ are insensitive to 
this inner extrapolation.

In Table 1 we list the ellipticals for which $r_{ex}$ can 
be accurately determined and the appropriate reference for 
$\Sigma_x(r)$ in the final column.
For several galaxies -- NGC 720, 4649, and 5044 -- 
$\Sigma_x(r)$ has been determined by more than one 
detector or method.
In the following discussion
we have determined $\Sigma_x(r)$ from the uppermost entry 
in Table 1,
but our results are not sensitive to this choice.
As can be seen from Table 1, the x-ray effective radius
$r_{ex}$ is relatively insensitive to the slightly
different band passes of 
the ROSAT and {\it Einstein} detectors.

For optical and x-ray luminosities and assumed 
distances in Table 1 
we have chosen values listed 
by Eskridge et al. (1995a).
We have made this choice for several reasons:
the wide scatter in $L_x$ is clearly seen in these data,
a uniform procedure has been used to determine $L_x$, 
and the substantial prior investment with these data by 
Eskridge, et al. (1995a, 1995b) in 
seeking statistical correlations.
As with the observed $L_x$ values, for group galaxies
we regard the hot gas as an intrinsic property of the 
central galaxy, not of the group or poor cluster.

In Figure 1a we show the scatter of 
these eleven well-resolved galaxies 
around the mean correlation line
$L_x \propto L_B^{2.0 \pm 0.2}$ 
determined with the E-M algorithm 
by Eskridge et al. (1995a) 
using all observed ellipticals 
in their sample (see their Figure 6a).
Most of our galaxies lie above the 
mean correlation since well-resolved ellipticals tend to be 
intrinsically more luminous.
We define residuals in Figure 1a 
as the vertical (logarithmic) distance 
of each galaxy from the mean correlation line.
In Figure 1b we plot these residuals against the dimensionless
ratio of the x-ray and optical sizes, $r_{ex}/r_e$.
A general correlation is apparent in Figure 1b, indicating
that the scatter in Figure 1a is in fact largely due to the 
variation of the relative sizes of the images.
The dashed line in Figure 1b is a linear fit
to the data assuming uniform errors.
The obvious correlation in Figure 1b is also real:
a Spearman rank-order test indicates a probability 
of only 0.02 that 
eleven randomly chosen points would produce a correlation 
of this strength.
This probability would be even smaller 
if the deviant outlier NGC 5044 were removed or if 
we had used the ASCA data of Fukazawa et al. (1996) to
determine $r_{ex}$.
We prefer the ROSAT data of David et al. (1994) for 
NGC 5044 since ROSAT has somewhat superior
spatial resolution.

A distance independent version of the
luminosity-size correlation is shown in Figure 2
which can be fit with least squares with slope 
$L_x/L_B \propto (r_{ex}/r_e)^{0.60 \pm 0.30}$.
Since $r_e$ and $L_B$ have a limited range
in the sample, this correlation is driven
primarily by $L_x$ and $r_{ex}$.
There is no obvious reason why NGC 5044 
does not participate in the correlation.
Because the x-ray emissivity varies as the square of the
plasma density, we expect quite generally
$L_x \propto \langle n_x \rangle^2 r_{ex}^3
\propto M_x^2 r_{ex}^{-3}$ where $M_x$ is the mass of
hot gas.
Therefore, the correlation
in Figure 2 implies a non-trivial
relationship between $\langle n_x \rangle$ or $M_x$ and
$r_{ex}$ that depends on some internal attribute of the
galaxies or their history.
Evidently, $r_{ex}/r_e$ is a ``second parameter'' in
the variation of $L_x$ with $L_B$, i.e.
$L_x \propto L_B^2 (r_{ex}/r_e)^{0.60 \pm 0.30}$.

The outer x-ray isophotes of these galaxies are influenced 
by the assumed background level which depends 
on their local environment.
To explore the sensitivity of 
the correlation in Figure 2 to the choice 
of background level, we arbitrarily truncated the 
images of the galaxies at $r_{cut} \le r_t$
and re-evaluated $r_{ex}$ and $L_x$ as functions of 
$r_{cut}$.
Figure 3 shows how rapidly $r_{ex}$ and $L_x$ approach
the values in Table 1 as $r_{cut} \rightarrow r_t$.
There is clearly a range in the degree of compactness  
among the x-ray images.
While most galaxies appear to be well resolved, 
values of $r_{ex}$ and $L_x$ for 
NGC 7619, 2563, 1399 and 4636 are still varying appreciably
with projected radius at $r = r_t$.
These galaxies also have the largest $r_{ex}$ in Table 1 and 
large $r_{ex}/r_e$ (Figure 2).
Several explanations are possible for 
this variety of convergence rates: it
may be a real effect demonstrating a lack of 
x-ray homology, it could arise from 
errors or inconsistencies in the 
adopted background levels or it could be an artifact 
of environmental influences on the hot gas 
including isophotal asymmetries.
NGC 4636, noticeably low in both plots in Figure 3 and 
known to be interacting with environmental gas,
has a markedly asymmetric x-ray image
(Trinchieri et al. 1994).
However, suppose that for some reason the assumed background 
was too large for these four galaxies 
and that $r_{ex}$ and $L_x$ actually continue 
to increase at $r > r_t$.
At $r = r_t$ the slopes $d \log L_x/ d \log r_{ex}$ for these
four galaxies are 0.6, 0.8, 0.7, and 0.5 respectively.
If the x-ray images for these galaxies
extend beyond $r_t$ in a smooth way, their points in
Figure 2 would move
toward the upper right, approximately along
the correlation.
Therfore, errors in choosing the local x-ray background
are unlikely to undermine the correlation in Figures 1b 
and 2 nor 
can they be large enough to change $L_x$ by almost 
two orders of magnitude, the full range represented 
in these figures.
The other seven ellipticals are better resolved and
less subject to background errors.
We conclude that the radial variation of x-ray 
emission differs among these ellipticals: some are 
compact (e.g. NGC 4649, 7626, 5044), others diffuse 
(e.g. NGC 4636, 1399).

\section{DISCUSSION} 

White and Sarazin (1991) examined many possible 
causes of the large dispersion in $L_x$.
Although they did not identify an intrinsic
property of ellipticals that correlated with residuals
in the $L_x,L_B$ diagram,
they did find that ellipticals with low $L_x/L_B$ 
have $\sim 50$\% more neighbor galaxies than those with 
high $L_x/L_B$. 
Assuming that the high incidence of neighbor galaxies 
indicates a higher galactic space density, 
White and Sarazin proposed that the scatter in $L_x$ may 
be related to some environmental effect such as 
ram pressure stripping or galactic mergers. 
Mackie \& Fabbiano (1997) also find a weak correlation 
of $L_x/L_B$ with the apparent local density of galaxies.
By comparison, the correlation we have found 
{\it does} involve a fundamental
intrinsic property of the galaxies, $r_{ex}/r_e$. 
Moreover, 
if our motivating conjecture of dynamical 
halo mass transfer is correct,
the observed disparity in $r_{ex}/r_e$ among ellipticals 
is likely to originate {\it not} in regions of very high density 
(i.e. rich clusters) where some of the bright ellipticals
currently reside, but in group environments where the 
ellipticals were originally formed 
and where mergers and 
tidal interactions were more efficient.
Low orbital velocities and hot gas densities in 
groups make ram stripping ($\propto \rho v^2$) less likely 
there but it could account for some lowering of $L_x$ 
in rich cluster environments (e.g. NGC 4406 in Virgo).

D'Ercole et al. (1989) and Ciotti et al. (1991) 
suggested that large variations in $L_x$
for given $L_B$ can occur if the interstellar 
gas in ellipticals is currently 
undergoing a dynamic readjustment from wind solutions
(low $L_x$) to subsonic flows (high $L_x$).
Initial galactic wind solutions occur in their models 
because of the large Type Ia supernova
rates assumed at early times.
The precise time that ellipticals
undergo the transition from wind to subsonic flow,
according to their proposal, 
depends on variations among other 
galactic parameters such as the mass and structure of
the dark halos or the stellar mass to light ratio,
resulting in the scatter in Figure 1a.
However, Loewenstein \& Mathews (1991) showed that
the current iron abundance in the hot interstellar gas 
would be 3 - 4 times solar it the Type Ia supernova 
rate had been large enough to drive winds at earlier times.
Since this is much larger than the mean iron abundance 
in ellipticals found by ASCA, $\lta$ 0.5 solar
(Loewenstein 1997),
a nearly synchronized dynamical 
readjustment of interstellar gas in 
ellipticals is unlikely to be responsible for the wide 
scatter in Figure 1a.

Our discovery of the $(r_{ex}/r_e)$-residual correlation
was motivated by the possibility that ellipticals exchange 
hot gas and dark matter during dynamical 
interactions in small groups,
but it is not easy to demonstrate conclusively 
that this explanation is correct.
The notion that halo material is disproportionally 
allocated to group ellipticals
is consistent with observations of 
elliptical-rich groups in which the x-ray emitting gas
is almost exclusively associated 
with the central group-dominant 
elliptical (Mulchaey et al. 1996; Mulchaey \& Zabludoff 1997).
Dynamical studies of evolving galaxy groups
(Merritt 1985; Bode et al. 1994; 
Garciagomez et al. 1996; Athanassoula et al 1997;
Dubinski 1997) describe how 
group-dominant ellipticals grow at the expense of
other member galaxies either by wholesale 
mergers or from tidally truncated halo debris.
Mergers were rapid initially since all group galaxies 
had massive halos, but later subsided as 
non-dominant galaxies began to interact primarily 
with the halo of the group-dominant galaxy;
this explains the large number of groups still 
present today (Zabludoff \& Mulchaey 1997).
The relative amount of dark matter and hot halo gas 
acquired by group-dominant 
ellipticals is stochastically variable.
The correlation in Figure 2 
may be a natural outcome of this group evolution 
in which $L_x$ and $r_{ex}$ 
in bright dominant ellipticals are enhanced while 
$L_x$ and $r_{ex}$ are reduced in donor ellipticals.
Since the dark halo and hot gas extend far beyond $r_e$,
relatively little stellar matter is tidally exchanged; 
donor and receiver ellipticals may have similar $L_B$.

\acknowledgments

We are pleased to thank Ann Zabludoff and John Mulchaey for 
a preview of their galaxy group observations and for 
informative discussions and emails.
Our work on the evolution of hot gas in ellipticals is supported by
NASA grant NAG 5-3060 for which we are very grateful. In addition
FB is supported 
in part by Grant ASI-95-RS-152 from the Agenzia Spaziale Italiana.




\makeatletter
\def\jnl@aj{AJ}
\ifx\revtex@jnl\jnl@aj\let\tablebreak=\nl\fi
\makeatother
\begin{deluxetable}{rrrrrrrrl}
\footnotesize
\tablewidth{7.0in}
\tablenum{1}
\tablecaption{OPTICAL AND X-RAY PROPERTIES OF ELLIPTICAL GALAXIES}
\tablehead{
\colhead{NGC} &
\colhead{$\log L_B$\tablenotemark{a}} &
\colhead{$\log L_x$\tablenotemark{a}} &
\colhead{$D$\tablenotemark{a}} &
\colhead{$r_e$\tablenotemark{b}} &
\colhead{$r_t$} &
\colhead{$r_{ex}$} &
\colhead{$r_{ex}/r_e$} &
\colhead{Ref. \& Det.~\tablenotemark{c}}  \nl
\colhead{} &
\colhead{(erg s$^{-1}$)} &
\colhead{(erg s$^{-1}$)} &
\colhead{(Mpc)} &
\colhead{(')} &
\colhead{(')} &
\colhead{(')} &
\colhead{} &
\colhead{}
}

\startdata
533  &   43.87 &  42.64 &  107.9 & 0.792 & 15.13 & 4.39 & 5.54 & (2)
~RP \cr
720  &   43.47 &  41.34 &  32.6  & 0.659 & 5.45 & 1.176 & 1.785 &
(8)~RP \cr
{}  &  {} &{}&{} & {}                    & 1.64 & 0.623 & 0.945 &
(8)~RH \cr
1399 &   43.46 &  42.28 &  27.2  & 0.706 & 25.1 & 11.21& 15.88 & (3)
~RP \cr
2563 &   43.70 &  42.12 &  96.1  & 0.467 & 12.82 & 5.55 & 11.88 & (2)
~RP \cr
4374 &   43.67 &  41.16 &  27.0  & 0.910 & 1.08 & 0.369 & 0.405 & (6)
~E \cr
4472 &   44.04 &  42.06 &  27.0  & 1.733 & 31.5 & 4.98 & 2.87 & (4)
~RP,RH \cr
4636 &   43.58 &  41.99 &  27.3  & 1.680 & 17.7 & 6.35 & 3.78 & (5)
~RP \cr
4649 &   43.83 &  41.61 &  27.0  & 1.227 & 9.42 & 0.827 & 0.674
& (2)~RP~\tablenotemark{d} \cr
{} & {}  &{} & {} &{} & 6.48 & 0.718 & 0.585
& (2)~RP~\tablenotemark{e}
\cr
&  {}  &  {} &{}  &{}  & 6.75 & 1.274 & 1.039 & (6)
~E \cr
5044 &   43.75 &  43.17 &  62.5  & 1.31 & 21.42 & 2.62 & 2.00 & (1)
~RP \cr
{} &   {} &  {} &  {}  & {} & 24.77 & 5.15 & 3.91  & (7)
~A \cr
7619 &   43.79   &42.00   &  75.0  & 0.536 & 22.8 & 6.75 & 12.60 & (2)
~RP \cr
7626 &   43.70 &  41.50 &  68.3  & 0.629 & 3.77 & 0.746 & 1.186 & (2)
~RP \cr

\enddata
\tablenotetext{a}{From Eskridge, Fabbiano, \& Kim (1995a).}

\tablenotetext{b}{Effective radii from Faber et al. (1989).}
\vskip.05in
\tablenotetext{c}{References and detectors
for $\Sigma_x$ and $r_{ex}$:
(1) David, Jones, Forman, \& Daines (1994),
(2) Trinchieri, Fabbiano, \& Kim (1997),
(3) Jones et al. (1997),
(4) Irwin \& Sarazin (1996),
(5) Trinchieri et al. (1994),
(6) Fabbiano, Kim \& Trinchieri (1992),
(7) Fukazawa et al. (1996),
(8) Buote \& Canizares (1996). Detectors: RP = ROSAT PSPC, RH = ROSAT
HRI, E = {\it Einstein} IPC and HRI, A = ASCA.
}
\tablenotetext{d}{Background normalization ``b''.}
\tablenotetext{e}{Background normalization ``a''.}
\end{deluxetable}


\vskip1.in
\figcaption[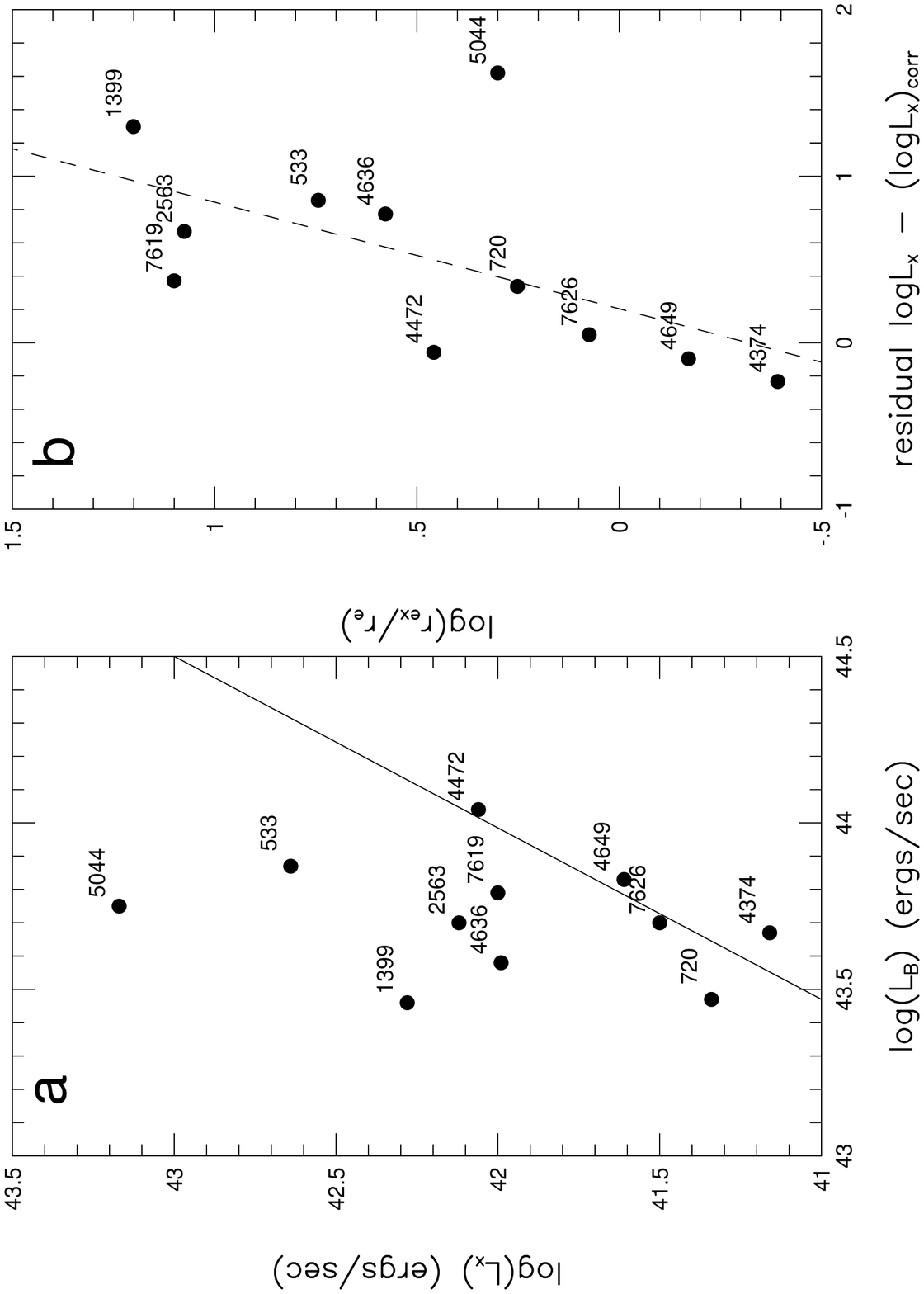]{(a) Location in the $L_x,L_B$ plane
of eleven ellipticals having well-resolved x-ray images.
The line shows the mean
correlation of all ellipticals observed with {\it Einstein}.
(b) Ratio of x-ray and optical effective radii
plotted against vertical residuals from the correlation
line in Fig 1a. The dashed line is a linear fit to the data.
 \label{fig1}}

\figcaption[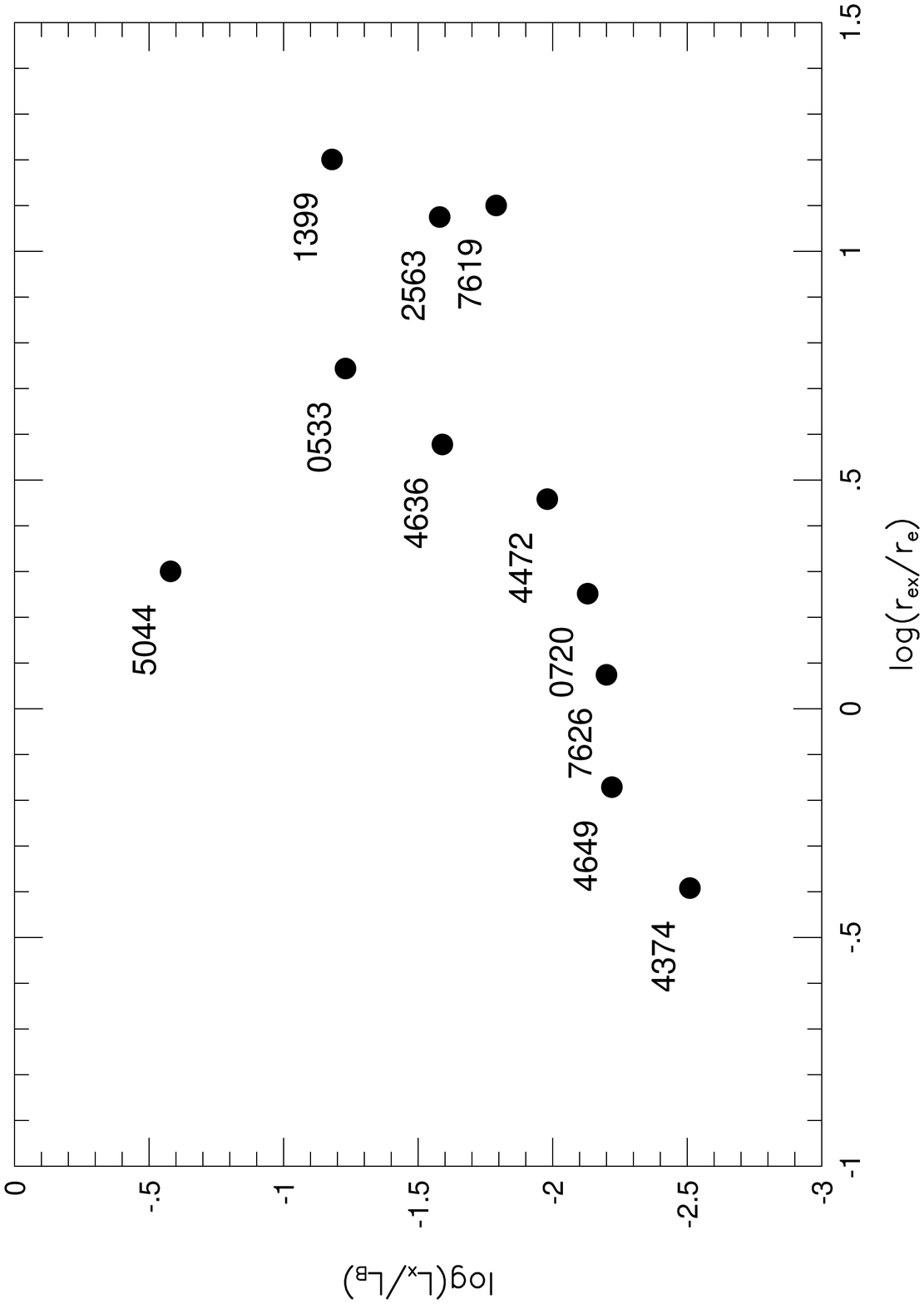]{Distance-independent plot
of $L_x/L_B$ against $r_{ex}/r_e$.\label{fig2}}

\figcaption[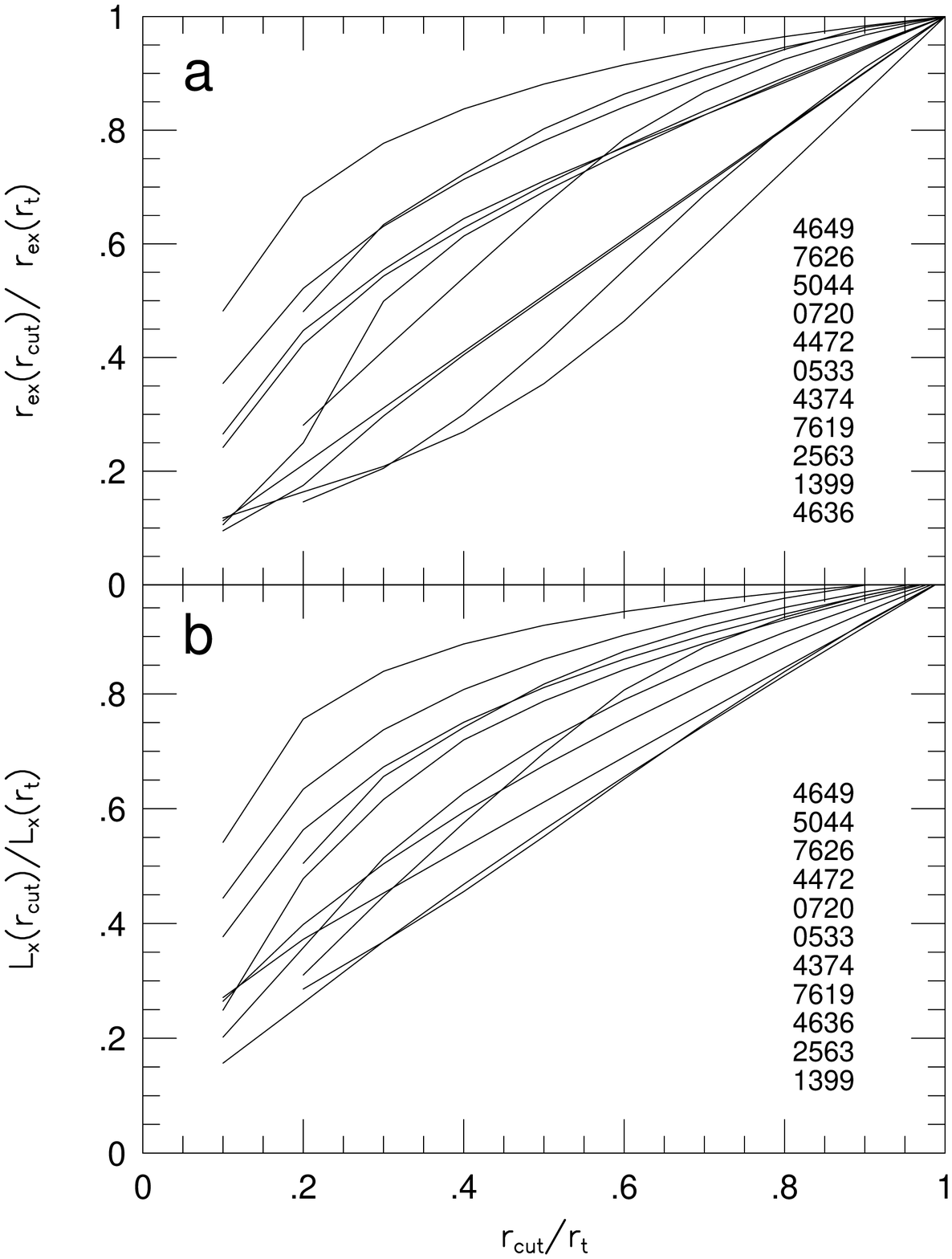]{(a) Variation of the
effective x-ray radius $r_{ex}$ as the abbreviated
image radius $r_{cut}$ increases toward the
outer radius of the x-ray image $r_t$;
(b) Variation of $L_x$ with $r_{cut}/r_t$.
The galaxy lists in both plots are in descending
order of $r_{ex}$ or $L_x$ evaluated at $r_{cut}/r_t = 0.5$.
\label{fig3}}


\begin{references}
\reference{head1988} Athanassoula, E., Makino, J., \& Bosma, A.
1997, MNRAS 286, 825
\reference{head1988} Brighenti, F. \& Mathews, W. G. 1997, ApJ
(in press) (astro-ph/9710199)
\reference{head1988} Bode, P. W., Berrington, R. C., Cohn, H. N. \&
Lugger, P. M. 1994, ApJ 433, 479
\reference{head1988} Buote, D. A. \& Canizares, C. R. 1996, ApJ 468,
184
\reference{head1988} Ciotti, L., D'Ercole, A., Pellegrini, S. \&
Renzini, A. 1991, ApJ 376, 380
\reference{head1988} David, L. P., Jones, C., Forman, W. \& Daines S.
1994, ApJ 428, 544
\reference{head1988} D'Ercole, A., Renzini, A., Ciotti, L, \&
Pellegrini, S. 1989, ApJ 341, L9
\reference{head1988} Dubinski, J. 1997, preprint (astro-ph/9709102)
\reference{head1988} Eskridge, P. B., Fabbiano, G., \& Kim, D-W.
1995a,
ApJS, 97, 141
\reference{head1988} Eskridge, P. B., Fabbiano, G., \& Kim, D-W.
1995b,
ApJ 442, 523
\reference{head1988} Fabbiano, G., Kim, D-W., \& Trinchieri, G. 1992,
ApJS 80, 531
\reference{head1988} Faber, S. M., Wegner, G., Burstein, D., Davies,
R. L., Dressler, A., Lynden-Bell, D., \& Terlevich, R. J. 
ApJS, 69, 763
\reference{head1988} Fukazawa, Y. et al. 1996, PASJ, 48, 395
\reference{head1988} Garciagomez, C., Athanassoula, E. \& Garijo, A.
1996, A\&A 313, 363
\reference{head1988} Irwin, J. A. \& Sarazin, C. L. 1996, ApJ 471,
683
\reference{head1988} Jones, C., Stern, C., Forman, W., Breen, J,
David, L. P., \& Tucker, W. 1997, ApJ 482, 143
\reference{head1988} Loewenstein, M. \& Mathews, W. G. 1991, ApJ 373,
445
\reference{head1988} Loewenstein, M. 1997,
{Galactic and Cluster Cooling
Flows}, ed. N. Soker, ASP Conf. Proc. 115, 100
\reference{head1988} Mackie, G. \& Fabbiano, G. 1997, 
{\it The Second Stromlo
Symposium: The Nature of Elliptical Galaxies} eds.
M. Arnaboldi, G. S. Da Costa \& P. Saha, ASP Conf. Proc. 116, 401
\reference{head1988} Merritt, D. 1985, ApJ 289, 18
\reference{head1988} Mulchaey, J. S., Davis, D. S., Mushotzky, R. F.
\& Burstein, D. 1996, ApJ 456, 80
\reference{head1988} Mulchaey, J. S. \& Zabludoff, A. L. 1997, ApJ
(preprint)
\reference{head1988} Trinchieri, G., Kim, D-W, Fabbiano, G.
Canizares, C. R. 1994, ApJ 428, 555
\reference{head1988} Trinchieri, G., Fabbiano, G., \& Kim, D-W. 1997,
A\&A, 318, 361
\reference{head1988} Tsai, J. C. \& Mathews, W. G. 1995, ApJ 448, 84
\reference{head1988} White, R. E., III, \& Sarazin, C. L. 1991, ApJ
367, 476
\reference{head1988} Zabludoff, A. I. \& Mulchaey, J. S. 1997, ApJ
(in press)
\end{references}
\end{document}